 \documentclass{aastex}
 \usepackage{emulateapj5}

\newcommand{\be}{\begin{equation}}
\newcommand{\ee}{\end{equation}}
\newcommand{\bdm}{\begin{displaymath}}
\newcommand{\edm}{\end{displaymath}}

\def\dmf{\dot{\mathfrak{M}}}

\begin{document}

\title{On a site of X-ray emission in AE~Aquarii}

  \author{Nazar R. Ikhsanov\altaffilmark{1}}

\affil{Institute of Astronomy, University of Cambridge, Madingley
Road, CB3~0HA, UK}

   \altaffiltext{1}{Central Astronomical Observatory, Russian Academy
   of Sciences, 65-1 Pulkovo, 196140 St.\,Petersburg, Russia}

\begin{abstract}
An analysis of recently reported results of XMM-Newton observations
of AE~Aqr within a hypothesis that the detected X-ray source is
located inside the Roche lobe of the white dwarf is presented. I
show this hypothesis to be inconsistent with the currently adopted
model of mass-transfer in the system. Possible solutions of this
problem are briefly discussed.
\end{abstract}


\section{Introduction}

AE~Aqr is a close binary system with an orbital period 9.88\,h, mass
ratio 0.58--0.89, and an inclination $i \simeq 55^{\degr} \pm
7^{\degr}$ (Welsh, Horne, \& Gomer 1995). The distance to the system
is about 100\,pc \citep{Friedjung-1997}. The degenerate companion is
a magnetized white dwarf rotating with a period of $P_{\rm s} =
33\,P_{33}$\,s and braking with a rate of $\dot{P}_0 = 5.64\times
10^{-14}\,{\rm s\,s^{-1}}$, which implies the spin-down power of
\citep{de-Jager-etal-1994,Welsh-1999},
   \be\label{lsd}
L_{\rm sd} \cong 6 \times 10^{33}\  P_{33}^{-3}\ I_{50}\
(\dot{P}/\dot{P}_0)\ {\rm erg\,s^{-1}},
  \ee
Here $I_{50}$ is the moment of inertia of the white dwarf expressed
in units of $10^{50}\,{\rm g\,cm^2}$.

The normal companion is a K3--K5 red dwarf which overflows its Roche
lobe and loses material through the L1 point towards the white
dwarf. This material manifests itself in a form of the optical/UV
continuum and emission lines. It is neither accreted onto the
surface of the white dwarf nor stored in a disk around its
magnetosphere. Instead, it interacts with the magnetic field of the
white dwarf via a drag term \citep{King-1993} and is leaving the
system without forming a disk (Wynn, King, \& Horne 1997; Welsh,
Horne, \& Gomer 1998; Ikhsanov, Neustroev, \& Beskrovnaya 2004).

The system X-ray emission has recently been studied with XMM-Newton
(RGS, EPIC) by Itoh, Ishida, \& Kunieda (2005). As they have
reported, the number density of plasma responsible for the detected
X-rays is $n_{\rm x}\sim 10^{11}\,{\rm cm^{-3}}$ and the linear
scale of the source is $\ell_{\rm x} \sim (2-3) \times 10^{10}$\,cm.
The observed spectrum has been well fitted using the 4-temperature
(0.14, 0.59, 1.4 and 4.6\,keV) VMEKAL model \citep{Mewe-etal-1995}.
The analysis of the centroids of N and O\,Ly$\alpha$ lines shows no
evidence for any significant orbital Doppler modulation. The widths
of these lines are close to $1000\,{\rm km\,s^{-1}}$.

Analyzing these results \citet{Itoh-etal-2005} have discarded a
possibility that the detected X-rays are emitted from the surface of
the white dwarf indicating that $n_{\rm x}$ is a few orders of
magnitude smaller than corresponding conventional estimates in the
post-shock accretion column and $\ell_{\rm x}$ exceeds the radius of
the white dwarf by almost two orders of magnitude. Instead, they
have suggested a hypothesis in which the detected emission is
associated with a heating and expansion of the material streaming
through Roche lobe of the white dwarf. According to their scenario
the heating occurs at an adiabatic shock located at a distance of
$\sim 10^{10}$\,cm from the white dwarf and the observed emission is
powered by the gravitational energy of the streaming material.

In this letter I show that the above mentioned hypothesis is
inconsistent with the currently adopted model of the mass transfer
in the system and thus, a question about a location of the source of
X-ray emission in AE~Aqr remains open.

   \section{Can the X-ray source be associated with an expanded
   stream\,?}\label{2}

Within the hypothesis suggested by \citet{Itoh-etal-2005} the flow
of hot material responsible for the observed X-rays is feeded by the
stream flowing into the Roche lobe of the white dwarf through the L1
point. The numerical simulations of H$\alpha$ Doppler tomogram of
AE~Aqr \citep{Wynn-etal-1997,Welsh-etal-1998,Ikhsanov-etal-2004}
have shown that a distance to which the stream could approach the
white dwarf, $r_{\rm min}$, significantly exceeds its corotation
radius $r_{\rm cor} \simeq 1.5 \times 10^9\ M_{0.8}^{1/3}\
P_{33}^{2/3}\ {\rm cm}$. This distance, in the general case, is
limited to $r_{\rm min} \ga \max[r_0, r_{\rm A}]$, where $r_{\rm A}$
is the Alfv\'en radius of the white dwarf, and
   \be\label{r0}
r_0 \simeq 10^{10} \left(\frac{q}{0.64}\right)^{ - 0.464}
 \left(\frac{a}{1.8 \times 10^{11}\,{\rm cm}}\right)\ {\rm cm}
   \ee
is a distance to which the material could approach the white dwarf
if its angular momentum along the whole trajectory remains constant
\citep[see e.g. Eq.~2.14 in][]{Warner-1995}. Here $q$ and $a$ are
the mass ratio and orbital separation of the system components.

The radial distance $r_0$ represents a minimum possible distance to
which the material flowing through the L1 point can approach the
white dwarf. The value of this parameter does not depend on either
the magnetic field strength of the white dwarf or the structure of
the inflowing material and is only based on the angular momentum
conservation law. For the material to come closer to the white dwarf
its angular momentum must be reduced. However a question about the
mechanism which could be responsible for such a reduction in the
case of AE~Aqr remains open. Indeed, due to a lack of a disk the
canonical viscosity model of the angular momentum transport is not
applicable. On the other hand, the interaction between the stream
and the magnetic field at a distance $r_0$ tends to increase the
angular momentum of the material since the velocity of field lines,
$\Omega r$, significantly exceeds the velocity of the material,
which is limited to the free-fall velocity, $V_{\rm ff} =
\sqrt{2GM_{\rm wd}/r}$.

A quantitative analysis of the mass-transfer process in AE~Aqr has
been first presented by \citet{Wynn-etal-1997}. As they have shown,
the H$\alpha$ Doppler tomogram of the system can be reproduced
assuming that the stream is inhomogeneous (a sequence of large
diamagnetic blobs) and interacts with the magnetic field of the
white dwarf via a drag term. The efficiency of this interaction is
$\propto r^{-n}$, where $n \geq 2$. That is why, the strongest
interaction between the blobs and the magnetic field occurs at their
closest approach to the white dwarf. The initial radius of the blobs
at $r_{\rm min}$ is
 \be\label{lb}
l_{\rm b} \simeq 10^9 \,Q_{19}^{1/2}\ {\rm cm},
 \ee
where $Q_{19}$ is the effective cross-section of the mass transfer
throat at the L1 point expressed in units of $10^{19}\,{\rm cm^2}$.
The initial sound speed and number density of the material entrained
in the blobs are, respectively
 \be
c_{\rm s} \simeq  10^6\ T_4^{1/2}\ {\rm cm\,s^{-1}},
 \ee
and
 \be\label{ne}
n_{\rm b} \simeq 10^{14}\ \dmf_{17}\ M_{0.8}^{-1/2}\ r_{10}^{1/2}
l_9^{-2}\ {\rm cm^{-3}}.
 \ee
Here $l_9=l_{\rm b}/10^9$\,cm, $r_{10}=r_{\rm min}/10^{10}$\,cm, and
$T_4$ is the mean temperature of the flowing material expressed in
units of $10^4$\,K
\citep[see][]{Beskrovnaya-etal-1996,Eracleous-Horne-1996}.

Let us now consider a scenario in which the hot flow forms at
$r_{\rm min}$ as a result of strong interaction between the blobs
and the magnetic field (independently of the mechanism of this
interaction). This implies a heating of the blobs (or their surface
layers) and their expansion up to a size of $\ell_{\rm x}$. As the
gas expands its density decreases by a factor of $\ell_{\rm
x}^3/(l_{\rm b}^2 \Delta s)$, where $\Delta s$ is the thickness of
the heated layer. This indicates that the initial number density of
material in the layer should satisfy the following condition
 \be\label{n0}
n_0 \ga n_{\rm x} \ell_{\rm x}/ (l_{\rm b}^2 \Delta s).
 \ee
Otherwise, the density of the hot flow would be significantly
smaller than that evaluated by \citet{Itoh-etal-2005}. Combining
Eqs.~(\ref{ne}) and (\ref{n0}), and setting $n_0 = n_{\rm b}$ one
finds
 \be
\Delta s \ga 8 \times 10^8\ \dmf_{17}^{-1}\ M_{0.8}^{1/2}\
r_{10}^{-1/2}\ \ell_{10}^3 \left(\frac{n_{\rm x}}{10^{11}\,{\rm
cm^{-3}}}\right) {\rm cm}.
 \ee
Thus, for the considered scenario to realize the entire blobs at a
distance $r_{\rm min}$ should be heated to a temperature of 5\,keV.

However, as soon as the temperature of the blob reaches 5\,keV, its
X-ray luminosity proves to be
 \be
L_{\rm x,sl} \simeq 6 \times 10^{32}\ T_{7.7}^{1/2}\ n_{14}^2\
l_9^3\ {\rm erg\,s^{-1}},
 \ee
i.e. a factor of 60 larger that the X-ray luminosity of the system
($\simeq 10^{31}\,{\rm erg\,s^{-1}}$, see e.g. Clayton \& Osborne
1995; Choi, Dotani, \& Agrawal 1999). Here $n_{14}=n_{\rm
b}/10^{14}\,{\rm cm^{-3}}$.

Furthermore, it remains unclear how the temperature of the blob
could reach the value of 5\,keV. Indeed, for the heating of the
blobs to occur the characteristic time of the heating process should
be smaller than the cooling time. But the bremsstrahlung cooling
time of the blobs,
 \be\label{tbr}
t_{\rm br} \simeq 15\ T_{7.7}^{1/2}\ n_{14}^{-1}\ {\rm s},
 \ee
is significantly smaller than the free-fall time at $r_{\rm min}$
($t_{\rm ff} \simeq 70\ r_{10}^{3/2} M_{0.8}^{-1/2}$\,s) and
correspondingly, the characteristic time of the drag interaction,
which according to \citet{Wynn-etal-1997} is limited to $t_{\rm
drag} > t_{\rm ff}$. The value of $t_{\rm br}$ is even smaller than
the turbulent diffusion time evaluated by
\citet{Meintjes-Venter-2005}. Thus, the assumption about a
significant heating of the entire blobs at the radius $r_{\rm min}$
within any currently considered mechanisms of interaction between
the blobs and the magnetic field of the white dwarf is not valid.
This means, that formation of the hot flow with the number density
$\sim 10^{11}\,{\rm cm^{-3}}$, linear scale $\ga 10^{10}$\,cm, and
temperature $\sim 5$\,keV inside the Roche lobe of the white dwarf
is impossible within the currently adopted model of the
mass-transfer. Instead, the source associated with the hot surface
layers of the blobs within this model is expected to be dense, $\sim
n_{\rm b}$, and compact, $\Delta s \la 3 \times 10^4\ n_{14}^{-2}\
T_{7.7}^{-1/2}$\,cm.

   \section{Discussion}

The problem posed in the previous section suggests that either our
current view on the mass-transfer picture in AE~Aqr is incomplete or
a significant part of detected X-rays is generated in a region
situated beyond the Roche lobes of the system components, or both.
Here I briefly address these two possibilities.

The situation can be partly improved if one invokes an assumption
that the mass transfer through the Roche lobe of the white dwarf
operates via more than one component. In particular, a formation of
the X-ray source with the required parameters inside the Roche lobe
of the white dwarf could be expected if the blobs were surrounded by
a medium of a linear size $\sim 10^{10}$\,cm and mean number density
$\sim 10^{11}\,{\rm cm^{-3}}$. The required rate of mass transfer by
this additional flow is $\sim 5 \times 10^{15}\,{\rm g\,s^{-1}}$,
i.e. significantly (by more than an order of magnitude) smaller than
the rate of mass transferred by the blobs. The heating of the flow,
according to Eq.~(\ref{tbr}), can be expected even if its
interaction with the magnetic field is gentle. In particular, the
cooling time of this flow would be larger than the time of heating
governed by the drag interaction \citep[see Eq.~1
in][]{Wynn-etal-1997}. The temperature of the flow, in the first
approximation, can be limited by equating the thermal pressure of
the material, $(1/2) n_{\rm x} m_{\rm p} c_{\rm s}^2$, with the
energy density of the external magnetic field, $B^2/8 \pi$:
 \be
 T \la 5.8 \times 10^7\ B_2^2\ \left(\frac{n_{\rm x}}{10^{11}\,{\rm
cm^{-3}}}\right)^{-1}\ {\rm K}.
 \ee
Here $B_2$ is the strength of the external magnetic field expressed
in units of $10^2$\,G.

It should be noted that both the density and size of the additional
flow component significantly exceed the corresponding parameters of
the interblob material considered by \citet{Wynn-etal-1997}. The hot
flow under these conditions screens the blobs from the
magnetospheric field making the drag interaction between them
ineffective. However, the suppression of the drag-driven propeller
action does not occur at the closest approach of the flow to the
white dwarf since the blobs and the hot flow at this distance follow
different trajectories. Indeed, being more dense the blobs would
significantly deeper penetrate into the magnetosphere of the white
dwarf \citep[for a discussion see
e.g.][]{Welsh-etal-1998,Ikhsanov-etal-2004}. This allows an ample
room for a direct interaction between the blobs and the magnetic
field of the white dwarf at the distance $r_0$, i.e. just at the
point where the efficiency of the drag interaction reaches its
maximum value.

At the same time, the origin of this additional component within the
currently adopted model of the system remains uncertain. An
incorporation of the additional component into the mass transfer
scheme implies that either its temperature at the L1 point exceeds
$2 \times 10^5$\,K (in this case a formation of the geometrically
thick flow at $r_{\rm min}$ can be treated in terms of its thermal
expansion), or the effective cross-section of the mass-transfer
throat at the L1 point is significantly larger than its canonical
value. A reason why these possibilities cannot be simply discarded
is that the normal component of AE~Aqr is partly (by almost a half
of its radius) situated inside the light cylinder of the white
dwarf. This situation is unique for all currently known close
binaries and the mass-transfer process which could be realized under
these conditions is poorly understood so far.

Another problem, which has to be addressed within the above
presented scenario, is a lack of any significant orbital Doppler
modulation of the N and O\,Ly$\alpha$ lines \citep{Itoh-etal-2005}.
The flow effectively contributes to the X-ray flux of the system
only on a time scale of its expansion,
 \be
t_{\rm exp} \simeq 160\ \ell_{10}\ T_{7.7}^{-1/2}\ {\rm s},
 \ee
and therefore, would appear as a local source. Indeed, the
temperature and density of an adiabatically expanding fully ionized
gas decrease as $T \propto l^{-2}$, and $n \propto l^{-3}$ and the
luminosity of the flow decreases on a time scale of $t_{\rm exp}$ by
more than an order of magnitude. This argues strongly against an
assumption that the hot gas fills the Roche lobe of the white dwarf
and spreads around the system \citep[as it has been suggested
by][]{Itoh-etal-2005}. Under these conditions, however, a modulation
of the flow parameters with the system orbital motion would be
expected and the lack of success in searching for this modulation
can hardly be interpreted in a simple way.

A modification of the currently adopted mass transfer model is not
required if the source of X-ray emission has a multi-component
nature. In particular, one can envisage a situation in which a part
of the detected X-rays are generated outside the system. The only
available energy source in this case is the spin-down power of the
white dwarf (the accretion power in this case can obviously be
excluded). Can this energy be converted into a 5\,keV emission in a
region situated outside the system\,?

The answer to this question depends on the ratio of the spin-down
power channeled into the mass outflow and particle acceleration
respectively, which is a matter of serious debate at present. It is
rather negative if the spin-down power is converted mainly into the
kinetic energy of the outflowing material. Indeed, the velocity
dispersion of the blobs leaving the system is limited to $\Delta V
\la 300\,{\rm km\,s^{-1}}$
\citep{Wynn-etal-1997,Welsh-etal-1998,Ikhsanov-etal-2004}. This is a
factor of 4 smaller than both the thermal velocity of a plasma
heated up to the temperature 5\,keV and the velocity evaluated from
the analysis of the width of emission lines (see Introduction).

The answer might be, however, positive if the spin-down power were
released in a form of accelerated particles or/and MHD waves. As
recently shown by \citet{Antonicci-Gomez-de-Castro-2005}, the
temperature of X-ray photons emitted by dense ($n_{\rm e} \sim
10^{11}\,{\rm cm^{-3}}$) blobs illuminated by a shower of
high-energy electrons (0.03-1\,MeV) ranges within an interval
1--20\,keV. The spectrum of this radiation depends on the angle
between the line of sight and the velocity vector of electrons. In
particular, the spectrum of blobs situated between an observer and
the source of accelerated particles differs from the spectrum
emitted by blobs situated behind the particle source. Therefore, an
accelerator of particles surrounded by the blobs would appear as a
multi-temperature source of 1--20\,keV emission.

Let us check whether such a situation can be realized in AE~Aqr. The
radius of dense blobs ejected from the system by the propeller
action of the white dwarf increases due to their thermal expansion
and reaches the value of $\ell_{\rm x}$ on a time scale
 \be\label{tau}
\tau = \frac{\ell_x}{c_{\rm s}} \simeq 10^4\ \ell_{10}\ c_6^{-1}\
{\rm s}.
 \ee
Here the speed of sound, $c_6 = c_{\rm s}/10^6\,{\rm cm\,s^{-1}}$,
is normalized by taking into account that the heating of the dense
blobs inside the Roche lobe of the white dwarf is insignificant (see
Sect.\,\ref{2}). On this time scale the blobs moving with an average
velocity $V_{\rm out}$ leave the system to a distance of
 \be
r_0 \sim V_{\rm out} \tau \simeq 3 \times 10^{11}\ \ell_{10}\
c_6^{-1} \left(\frac{V_{\rm out}}{300\,{\rm km\,s^{-1}}}\right) {\rm
cm}
  \ee
and their number density decreases to
 \be
n_0 = n_{\rm b} \left(\frac{l_{\rm b}}{\ell_{\rm x}}\right)^3 \simeq
10^{11}\ l_9^3\ \ell_{10}^{-3} \left(\frac{n_{\rm b}}{10^{14}\,{\rm
cm^{-3}}}\right)\ {\rm cm^{-3}}
 \ee

The number of blobs ejected by the white dwarf during the time
$\tau$ is
 \be\label{ndif}
N \simeq  \frac{3\ \dmf\ \tau}{4 \pi n_{\rm b} m_{\rm p} l_{\rm
b}^3}.
 \ee

Combining Eqs.~(\ref{ne}), (\ref{tau}), and (\ref{ndif}) yields
 \be
N \simeq 10^3\ M_{0.8}^{1/2}\ c_6^{-1}\ \ell_{10}\ l_9^{-3}\
r_{10}^{-1/2}.
 \ee

Therefore, the stream of the expanded blobs will occupy an area of
 \be\label{A}
A \sim \pi \ell_{\rm x}^2 N \simeq 3 \times 10^{23}\
\left(\frac{\ell_{\rm x}}{10^{10}\,{\rm cm}}\right)^2
\left(\frac{N}{10^3}\right)\ {\rm cm^2},
 \ee
which constitutes $\sim 0.3$ of the total area of a sphere with a
radius of $r_0$. This indicates that 30\% of the energy released in
the system will be transferred through the stream of the expanded
blobs.

If now we assume that the spin-down power releases predominantly in
a form of accelerated particles \citep[see
e.g][]{Ikhsanov-1998,Meintjes-de-Jager-2000} one finds the total
flux of the energy transferred through the area occupied by the
stream of the expanded blobs as (see Eqs.~\ref{ne}, and
\ref{tau}--\ref{A})
 \be
L_1 = L_{\rm sd} \frac{A}{4 \pi r_0^2} \simeq 10^{33}\ {\rm
erg\,s^{-1}}\ \ell_{10} M_{0.8}^{1/2}\ c_6\ l_9^{-1}\ r_{10}^{-1/2}\
\times
 \ee
 \bdm
\left(\frac{V_{\rm out}}{300\,{\rm km\,s^{-1}}}\right)^{-2}
\left(\frac{L_{\rm sd}}{6 \times 10^{33}\,{\rm erg\,s^{-1}}}\right).
 \edm
Hence, a conversion of only 2\% of this energy into X-rays would be
enough for the luminosity of the corresponding source to be
comparable to the X-ray luminosity observed from the system.

The above estimates suggest that an interpretation of the X-ray
emission of AE~Aqr in terms of expanded blobs illuminated by a flux
of accelerated particles might be promising. A further analysis of
this scenario, however, requires information about the spectrum and
geometry of the wind of particles ejected by the white dwarf, which
is currently not available. Furthermore, a question about the region
of the wind formation also remains open. If this region \citep[as in
the case of spin-powered pulsars, see e.g.][]{Beskin-etal-1993} is
situated at the light cylinder, the surface of the normal companion
would also be partly affected by the wind. However, a contribution
of this additional source into the X-ray flux can unlikely be
significant since the area occupied by the normal companion at the
radius of the light cylinder is a factor of 20 smaller than A (see
Eq.~\ref{A}).

\acknowledgments

I thank James Pringle, Andrew Fabian and Christopher Mauche for
useful discussions and an anonymous referee for suggesting
improvements. I acknowledges the support of the European Commission
under the Marie Curie Incoming Fellowship Program. The work was
partly supported by Russian Foundation of Basic Research.

\end{document}